\documentclass[twocolumn,superscriptaddress,amsmath,amssymb,prc,aps]{revtex4-1}
\usepackage{graphicx}  
\usepackage{siunitx}  
\usepackage{epsfig} 
\usepackage{subfigure}
\usepackage{lineno}  
\usepackage{amsmath,amssymb}
\usepackage{comment}
\usepackage{multirow}
\usepackage{dcolumn}
\usepackage{bm}
\usepackage{setspace}
\usepackage{multibib}
\includecomment{printrobustnotes}
\includecomment{printallnotes}
\usepackage{xspace}
\usepackage{color}
\begin{document}

\title{An induced  annual modulation signature in COSINE-100 data by DAMA/LIBRA's analysis method}
\author{G.~Adhikari}
\affiliation{Department of Physics and Wright Laboratory, Yale University, New Haven, CT 06520, USA}
\author{N.~Carlin}
\affiliation{Physics Institute, University of S\~{a}o Paulo, 05508-090, S\~{a}o Paulo, Brazil}
\author{J.~J.~Choi}
\affiliation{Department of Physics and Astronomy, Seoul National University, Seoul 08826, Republic of Korea} 
\affiliation{Center for Underground Physics, Institute for Basic Science (IBS), Daejeon 34126, Republic of Korea}
\author{S.~Choi}
\affiliation{Department of Physics and Astronomy, Seoul National University, Seoul 08826, Republic of Korea} 
\author{A.~C.~Ezeribe}
\affiliation{Department of Physics and Astronomy, University of Sheffield, Sheffield S3 7RH, United Kingdom}
\author{L.~E.~Fran{\c c}a}
\affiliation{Physics Institute, University of S\~{a}o Paulo, 05508-090, S\~{a}o Paulo, Brazil}
\author{C.~Ha}
\affiliation{Department of Physics, Chung-Ang University, Seoul 06973, Republic of Korea}
\author{I.~S.~Hahn}
\affiliation{Department of Science Education, Ewha Womans University, Seoul 03760, Republic of Korea} 
\affiliation{Center for Exotic Nuclear Studies, Institute for Basic Science (IBS), Daejeon 34126, Republic of Korea}
\affiliation{IBS School, University of Science and Technology (UST), Daejeon 34113, Republic of Korea}
\author{S.~J.~Hollick}
\affiliation{Department of Physics and Wright Laboratory, Yale University, New Haven, CT 06520, USA}
\author{E.~J.~Jeon}
\affiliation{Center for Underground Physics, Institute for Basic Science (IBS), Daejeon 34126, Republic of Korea}
\author{J.~H.~Jo}
\affiliation{Department of Physics and Wright Laboratory, Yale University, New Haven, CT 06520, USA}
\author{H.~W.~Joo}
\affiliation{Department of Physics and Astronomy, Seoul National University, Seoul 08826, Republic of Korea} 
\author{W.~G.~Kang}
\affiliation{Center for Underground Physics, Institute for Basic Science (IBS), Daejeon 34126, Republic of Korea}
\author{M.~Kauer}
\affiliation{Department of Physics and Wisconsin IceCube Particle Astrophysics Center, University of Wisconsin-Madison, Madison, WI 53706, USA}
\author{B.~H.~Kim}
\affiliation{Center for Underground Physics, Institute for Basic Science (IBS), Daejeon 34126, Republic of Korea}
\author{H.~J.~Kim}
\affiliation{Department of Physics, Kyungpook National University, Daegu 41566, Republic of Korea}
\author{J.~Kim}
\affiliation{Department of Physics, Chung-Ang University, Seoul 06973, Republic of Korea}
\author{K.~W.~Kim}
\affiliation{Center for Underground Physics, Institute for Basic Science (IBS), Daejeon 34126, Republic of Korea}
\author{S.~H.~Kim}
\affiliation{Center for Underground Physics, Institute for Basic Science (IBS), Daejeon 34126, Republic of Korea}
\author{S.~K.~Kim}
\affiliation{Department of Physics and Astronomy, Seoul National University, Seoul 08826, Republic of Korea}
\author{W.~K.~Kim}
\affiliation{IBS School, University of Science and Technology (UST), Daejeon 34113, Republic of Korea}
\affiliation{Center for Underground Physics, Institute for Basic Science (IBS), Daejeon 34126, Republic of Korea}
\author{Y.~D.~Kim}
\affiliation{Center for Underground Physics, Institute for Basic Science (IBS), Daejeon 34126, Republic of Korea}
\affiliation{Department of Physics, Sejong University, Seoul 05006, Republic of Korea}
\affiliation{IBS School, University of Science and Technology (UST), Daejeon 34113, Republic of Korea}
\author{Y.~H.~Kim}
\affiliation{Center for Underground Physics, Institute for Basic Science (IBS), Daejeon 34126, Republic of Korea}
\affiliation{Korea Research Institute of Standards and Science, Daejeon 34113, Republic of Korea}
\affiliation{IBS School, University of Science and Technology (UST), Daejeon 34113, Republic of Korea}
\author{Y.~J.~Ko}
\affiliation{Center for Underground Physics, Institute for Basic Science (IBS), Daejeon 34126, Republic of Korea}
\author{D.~H.~Lee}
\affiliation{Department of Physics, Kyungpook National University, Daegu 41566, Republic of Korea}
\author{E.~K.~Lee}
\affiliation{Center for Underground Physics, Institute for Basic Science (IBS), Daejeon 34126, Republic of Korea}
\author{H.~Lee}
\affiliation{IBS School, University of Science and Technology (UST), Daejeon 34113, Republic of Korea}
\affiliation{Center for Underground Physics, Institute for Basic Science (IBS), Daejeon 34126, Republic of Korea}
\author{H.~S.~Lee}
\email{hyunsulee@ibs.re.kr}
\affiliation{Center for Underground Physics, Institute for Basic Science (IBS), Daejeon 34126, Republic of Korea}
\affiliation{IBS School, University of Science and Technology (UST), Daejeon 34113, Republic of Korea}
\author{H.~Y.~Lee}
\affiliation{Center for Underground Physics, Institute for Basic Science (IBS), Daejeon 34126, Republic of Korea}
\author{I.~S.~Lee}
\affiliation{Center for Underground Physics, Institute for Basic Science (IBS), Daejeon 34126, Republic of Korea}
\author{J.~Lee}
\affiliation{Center for Underground Physics, Institute for Basic Science (IBS), Daejeon 34126, Republic of Korea}
\author{J.~Y.~Lee}
\affiliation{Department of Physics, Kyungpook National University, Daegu 41566, Republic of Korea}
\author{M.~H.~Lee}
\affiliation{Center for Underground Physics, Institute for Basic Science (IBS), Daejeon 34126, Republic of Korea}
\affiliation{IBS School, University of Science and Technology (UST), Daejeon 34113, Republic of Korea}
\author{S.~H.~Lee}
\affiliation{IBS School, University of Science and Technology (UST), Daejeon 34113, Republic of Korea}
\affiliation{Center for Underground Physics, Institute for Basic Science (IBS), Daejeon 34126, Republic of Korea}
\author{S.~M.~Lee}
\affiliation{Department of Physics and Astronomy, Seoul National University, Seoul 08826, Republic of Korea} 
\author{Y.~J.~Lee}
\affiliation{Department of Physics, Chung-Ang University, Seoul 06973, Republic of Korea}
\author{D.~S.~Leonard}
\affiliation{Center for Underground Physics, Institute for Basic Science (IBS), Daejeon 34126, Republic of Korea}
\author{B.~B.~Manzato}
\affiliation{Physics Institute, University of S\~{a}o Paulo, 05508-090, S\~{a}o Paulo, Brazil}
\author{R.~H.~Maruyama}
\affiliation{Department of Physics and Wright Laboratory, Yale University, New Haven, CT 06520, USA}
\author{R.~J.~Neal}
\affiliation{Department of Physics and Astronomy, University of Sheffield, Sheffield S3 7RH, United Kingdom}
\author{J.~A.~Nikkel}
\affiliation{Department of Physics and Wright Laboratory, Yale University, New Haven, CT 06520, USA}
\author{S.~L.~Olsen}
\affiliation{Center for Underground Physics, Institute for Basic Science (IBS), Daejeon 34126, Republic of Korea}
\author{B.~J.~Park}
\affiliation{IBS School, University of Science and Technology (UST), Daejeon 34113, Republic of Korea}
\affiliation{Center for Underground Physics, Institute for Basic Science (IBS), Daejeon 34126, Republic of Korea}
\author{H.~K.~Park}
\affiliation{Department of Accelerator Science, Korea University, Sejong 30019, Republic of Korea}
\author{H.~S.~Park}
\affiliation{Korea Research Institute of Standards and Science, Daejeon 34113, Republic of Korea}
\author{K.~S.~Park}
\affiliation{Center for Underground Physics, Institute for Basic Science (IBS), Daejeon 34126, Republic of Korea}
\author{S.~D.~Park}
\affiliation{Department of Physics, Kyungpook National University, Daegu 41566, Republic of Korea}
\author{R.~L.~C.~Pitta}
\affiliation{Physics Institute, University of S\~{a}o Paulo, 05508-090, S\~{a}o Paulo, Brazil}
\author{H.~Prihtiadi}
\email{hafizhp@ibs.re.kr}
\affiliation{Center for Underground Physics, Institute for Basic Science (IBS), Daejeon 34126, Republic of Korea}
\author{S.~J.~Ra}
\affiliation{Center for Underground Physics, Institute for Basic Science (IBS), Daejeon 34126, Republic of Korea}
\author{C.~Rott}
\affiliation{Department of Physics, Sungkyunkwan University, Suwon 16419, Republic of Korea}
\affiliation{Department of Physics and Astronomy, University of Utah, Salt Lake City, UT 84112, USA}
\author{K.~A.~Shin}
\affiliation{Center for Underground Physics, Institute for Basic Science (IBS), Daejeon 34126, Republic of Korea}
\author{A.~Scarff}
\affiliation{Department of Physics and Astronomy, University of Sheffield, Sheffield S3 7RH, United Kingdom}
\author{N.~J.~C.~Spooner}
\affiliation{Department of Physics and Astronomy, University of Sheffield, Sheffield S3 7RH, United Kingdom}
\author{W.~G.~Thompson}
\affiliation{Department of Physics and Wright Laboratory, Yale University, New Haven, CT 06520, USA}
\author{L.~Yang}
\affiliation{Department of Physics, University of California San Diego, La Jolla, CA 92093, USA}
\author{G.~H.~Yu}
\affiliation{Department of Physics, Sungkyunkwan University, Suwon 16419, Republic of Korea}
\affiliation{Center for Underground Physics, Institute for Basic Science (IBS), Daejeon 34126, Republic of Korea}
\collaboration{COSINE-100 Collaboration}


\date{\today}

\begin{abstract} 
		The DAMA/LIBRA collaboration has reported the observation of an annual modulation in the event rate that has been attributed to dark matter interactions over the last two decades~\cite{Bernabei:2013xsa,Bernabei:2018yyw}. 
		However, even though tremendous efforts to detect similar dark matter interactions were pursued, no definitive evidence has been observed to corroborate the DAMA/LIBRA signal~\cite{Zyla:2020zbs,Billard:2021uyg}. 
		Many studies assuming various dark matter models have attempted to reconcile DAMA/LIBRA's modulation signals and null results from other experiments, however no clear conclusion can be drawn~\cite{Kang:2018zld,Herrero-Garcia:2018lga,Kang:2019uuj,COSINE-100:2021xqn}.  
		Apart from the dark matter hypothesis, several studies have examined the possibility that the modulation is induced by variations in their detector's environment or their specific analysis methods.
In particular, a recent study presents a possible cause of the annual modulation from an analysis method adopted by the DAMA/LIBRA experiment in which the observed annual modulation could be reproduced by a slowly varying time-dependent background~\cite{Buttazzo2020}. 
		Here, we study the COSINE-100 data using an analysis method similar to the one adopted by the DAMA/LIBRA experiment and observe a significant annual modulation, although the modulation phase is almost opposite to that of the DAMA/LIBRA data. Assuming the same background composition for COSINE-100 and DAMA/LIBRA, simulated experiments for the DAMA/LIBRA without dark matter signals also provide significant annual modulation with an amplitude similar to DAMA/LIBRA with opposite phase. Even though this observation does not explain the DAMA/LIBRA's results directly, this interesting phenomenon motivates deeper studies of the time-dependent DAMA/LIBRA background data.
\end{abstract}
\maketitle

		Several groups have attempted to develop experiments aiming at reproducing or refuting DAMA/LIBRA's results using the same NaI(Tl) target material~\cite{deSouza:2016fxg,Amare:2019jul,Antonello2019}. The COSINE-100 experiment~\cite{Adhikari:2017esn,Adhikari:2019off} is one of these that is currently operating with 106\,kg of low-background NaI(Tl) crystals at the Yangyang underground laboratory (Y2L). 
		Dark matter interpretations of the COSINE-100 data from the background spectra presented null observations~\cite{Adhikari:2018ljm,COSINE-100:2021xqn} that were inconsistent with an explanation of the DAMA/LIBRA signals as a spin-independent interaction between weakly interacting massive particles (WIMPs), the stringent candidate of the dark matter particle, and sodium or iodine nuclei in the specific context of the standard halo model. 
		Model independent searches of the annual modulation were reported~\cite{Adhikari:2019off,COSINE-100:2021zqh} but with insufficient statistics to corroborate the DAMA/LIBRA observation yet. Another experiment currently in operation, ANAIS-112, also reported the model-independent annual modulation search results that exhibited more than 2$\sigma$ tension with DAMA/LIBRA's signals~\cite{ANAIS2021} but has not yet made a final conclusion. 
		These model-independent searches considered time-dependent backgrounds based on the time-dependence of the rate induced by individual isotopes studied by cosmogenic activations~\cite{deSouza:2019hpk,Villar2018aab} and precise background modeling~\cite{cosinebg,Amare:2018ndh,cosinebg2}. 
		However, the DAMA/LIBRA experiment used the residual rate by subtracting an average rate in every one year cycle of data-taking roughly starting in September~\cite{Bernabei:2013xsa,Bernabei:2018yyw}. 
		If the background rate is not constant over time, this procedure can generate annually modulated event rates~\cite{Buttazzo2020,Messina_2020}.  Specifically, a slowly increasing event rate as a function of the time in the region of interest (ROI) can provide an annual modulation similar to the one observed by DAMA/LIBRA without dark matter signals as studied in literature~\cite{Buttazzo2020}. 
		To verify this phenomenon, it is interesting to apply DAMA/LIBRA's analysis technique to other experimental data.
		Here we analyze the COSINE-100 data in terms of the annual modulation, but applying the analysis methods adopted by DAMA/LIBRA. 
		
		The COSINE-100 experiment~\cite{Adhikari:2017esn} started physics operation in September 2016 at Y2L in South Korea with about 700~m of rock overburden. It utilizes eight low-background NaI(Tl) crystals arranged in a 4$\times$2 array, with a total target mass of 106\,kg. Each crystal is coupled to two photomultiplier tubes~(PMTs) to measure the amount of energy deposited in the crystal. The NaI(Tl) detectors are immersed in a 2,200 liter liquid scintillator, which allows for the identification and subsequent reduction of radioactive backgrounds observed by the crystal~\cite{Adhikari:2020asl}. The liquid scintillator is surrounded by copper, lead, and plastic scintillators to reduce the background contribution from external radiation as well as cosmic-ray muons~\cite{Prihtiadi:2017inr,Prihtiadi:2020yhz}.

		In contrast with the COSINE-100 detector, DAMA/LIBRA does not employ plastic- or liquid scintillator-based veto detectors. 
		We, therefore, do not use information from those detectors in this analysis. 
		In the ROI, PMT-induced noise events predominantly contribute to the single-hit physics data. They typically have fast decay times of less than 50\,ns compared with typical NaI(Tl) scintillation of about 250\,ns. 
		The DAMA/LIBRA experiment developed a parameter to discriminate the PMT-induced noise by using a ratio of fast charge between 0 and 50\,ns, X$_1$, and slow charge between 100 and 600\,ns, X$_2$, each defined from relative to a time in the rising edge of PMT's waveform. The event selection (ES) parameter based on X$_1$ and X$_2$ provided good separation of the PMT-induced noise as discussed in Appendix. 
		DAMA/LIBRA claimed that they could efficiently remove the PMT-induced noise events only using the ES parameter cut and achieve a 1\,keV energy threshold with almost no noise contamination~\cite{Bernabei:2020mon}.
		It should be noted that the 1\,keV energy threshold from the COSINE-100 data was achieved by a multivariable machine learning technique (COSINE-100 nominal event selection) that used multiple parameters including, but not limited to X$_1$ and X$_2$, mean decay time, charge asymmetry between two PMTs, and likelihood parameters for signal-like and noise-like templates~\cite{COSINE-100:2020wrv,COSINE-100:2021xqn}.

\begin{figure}[!htb]
  \begin{center}
    \includegraphics[width=1.0\columnwidth]{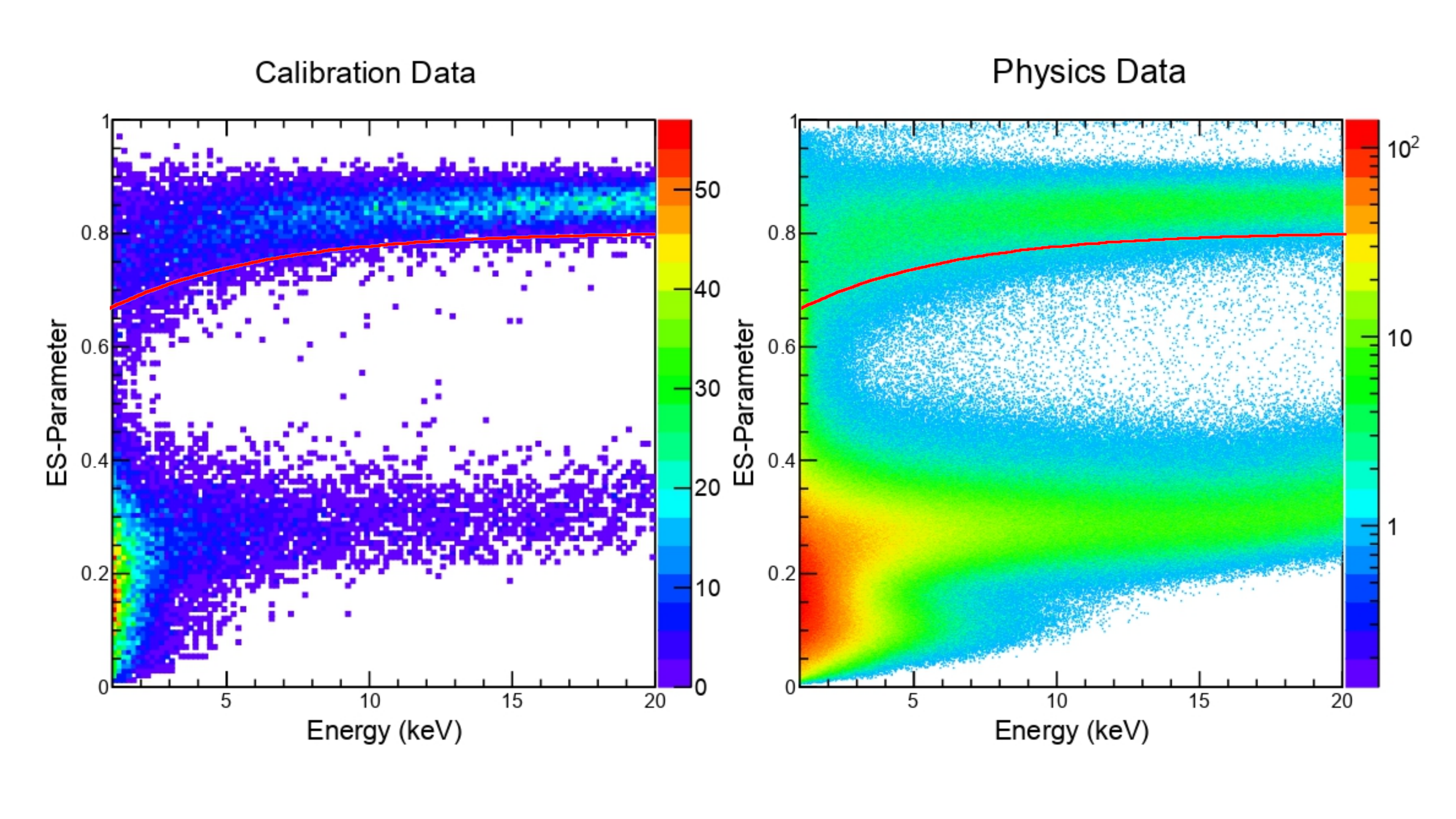} 
  \end{center}
  \caption{ 
    {\bf The ES parameter vs energy applied to data from COSINE-100.}
		The DAMA's ES parameters for the PMT-induced noise rejection as a function of energy are presented for the $^{60}$Co calibration (left) and three years COSINE-100 physics data (right). 
 Red solid-lines present selection criteria that provide selection efficiencies similar to the DAMA/LIBRA-phase2.
}
  \label{fig:eventselection}
\end{figure}

\begin{figure}[!htb]
  \begin{center}
    \includegraphics[width=1.0\columnwidth]{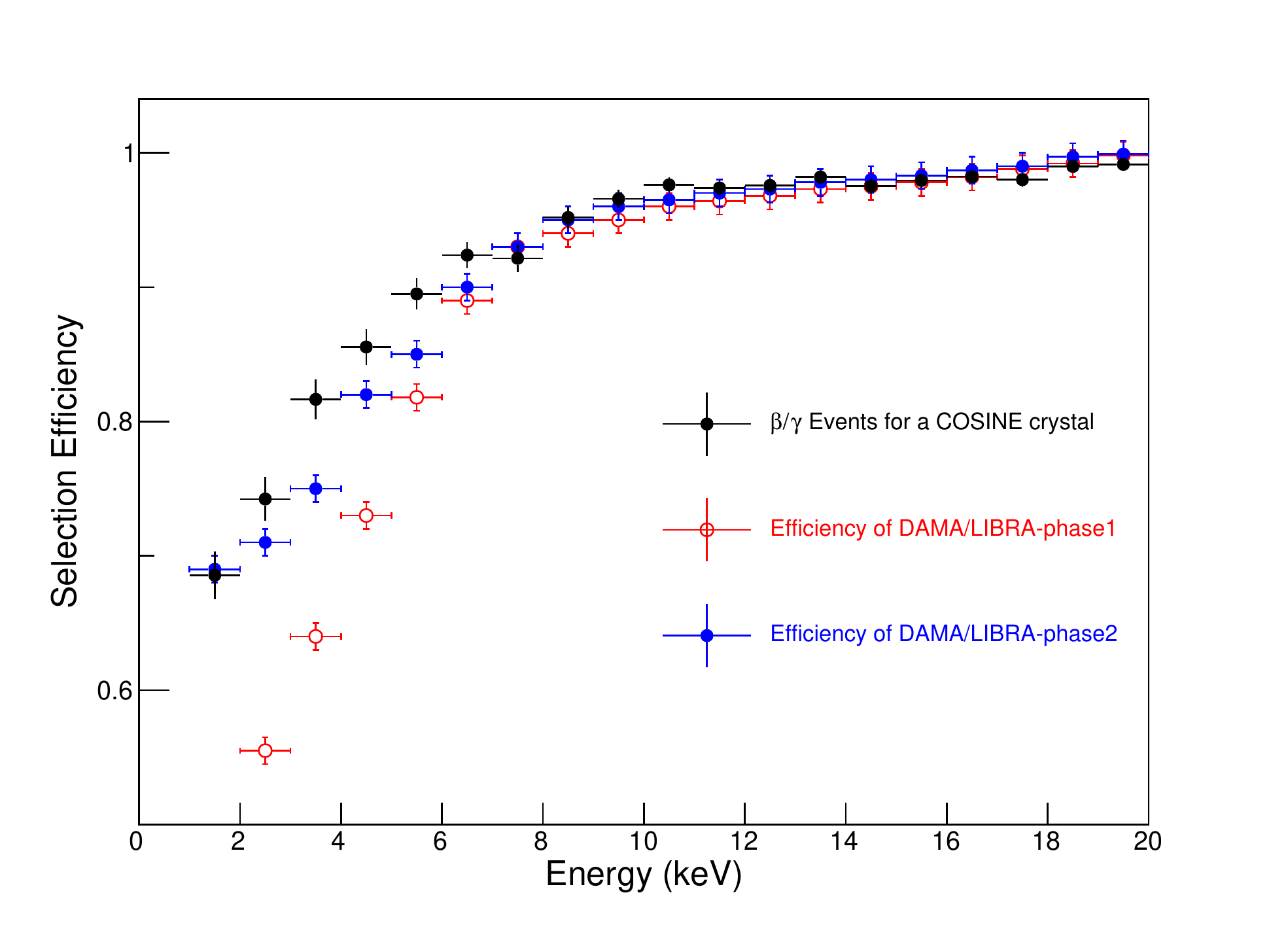} 
  \end{center}
  \caption{ 
    {\bf Event selection efficiency with the DAMA/LIBRA's event selection applied to COSINE-100 data. }
		Event selection efficiencies determined from the $^{60}$Co calibration data of the COSINE-100 crystal (crystal 6) (black points) are compared with those of DAMA/LIBRA-phase1 (red points) and DAMA/LIBRA-phase2 (blue points). 
		Here selection criteria shown in Fig.~\ref{fig:eventselection} are determined in order to provide efficiencies similar to the DAMA/LIBRA-phase2. 
}
  \label{fig:efficiency}
\end{figure}

We found slightly different selection efficiencies when we applied exactly the same criteria used by the DAMA/LIBRA for the ES parameter. 
Instead of the same value of the ES parameter cut, we choose values of the ES cut that result in a selection efficiency similar to the DAMA/LIBRA-phase2~\cite{Bernabei:2020mon}. 
We evaluate the selection efficiency with a $^{60}$Co calibration dataset, which has been used for the COSINE-100 data analyses~\cite{Adhikari:2018ljm,COSINE-100:2021xqn,Adhikari:2019off,COSINE-100:2021zqh}, although DAMA/LIBRA used $^{241}$Am calibration data. 
Our selection criteria developed with the $^{60}$Co calibration data and applied to the physics data are presented in Fig.~\ref{fig:eventselection}. 
The event selection efficiency compared with DAMA/LIBRA's efficiencies are shown in Fig.~\ref{fig:efficiency}.

\begin{figure}[!htb]
  \begin{center}
    \includegraphics[width=1.0\columnwidth]{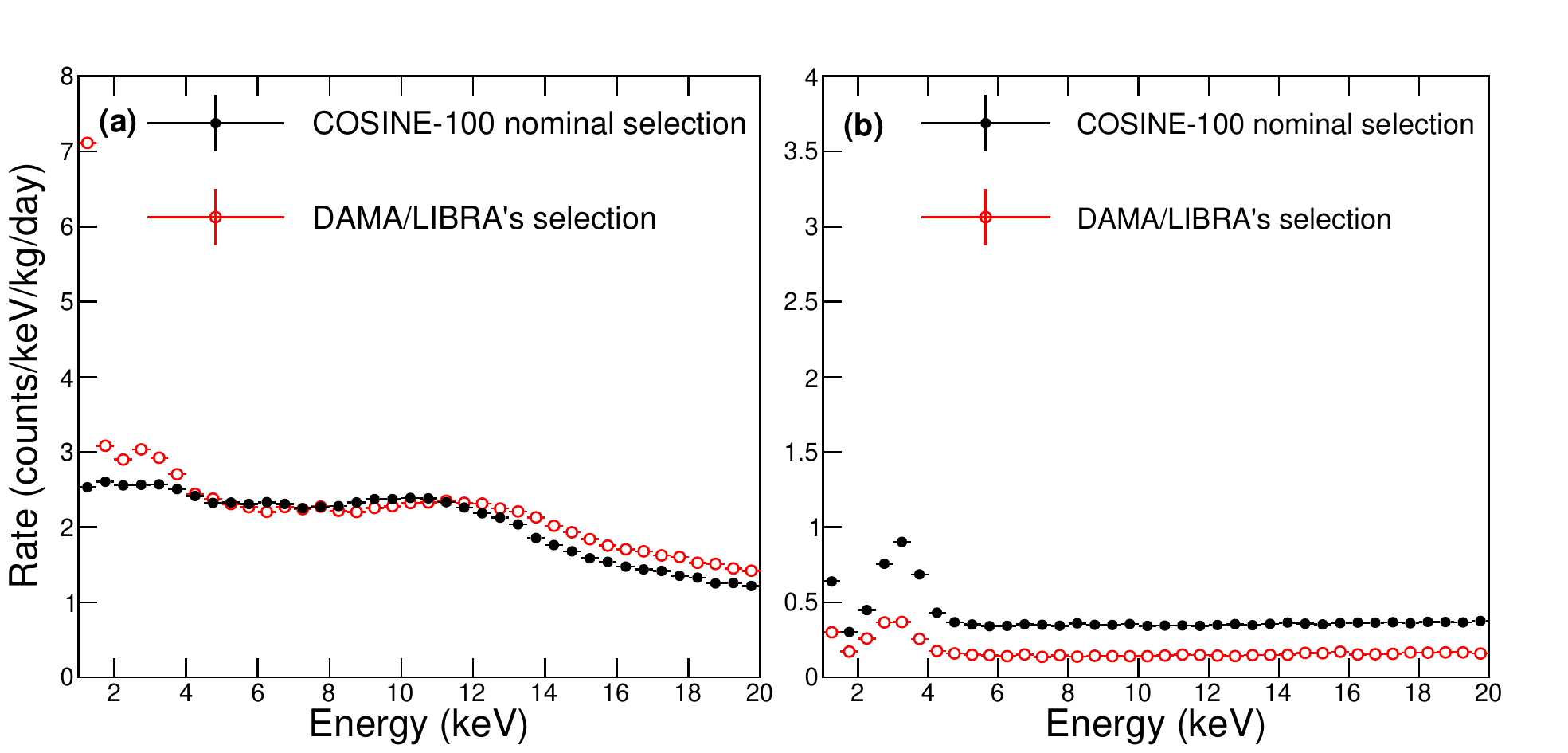} 
  \end{center}
  \caption{ 
    {\bf Low-energy spectra of a NaI(Tl) crystal in the  COSINE-100 experiment. }
		Energy spectra of one crystal (crystal 6) in COSINE-100 using nominal COSINE-100 event selection (black filled circles) and the DAMA/LIBRA's event selection (red open circles) are presented for the single-hit events (a) and multiple-hit events (b). Here, selection efficiencies are corrected for proper comparison. 
		Because we do not use the muon and LS detectors for the DAMA/LIBRA's event selection, the COSINE-100 nominal analysis obtains significantly larger numbers of the multiple-hit events. Due to the remnants of the PMT-induced events from the DAMA/LIBRA's event selection, significant excess of the event rate below 2\,keV in the single-hit events is observed.
}
\label{fig:energyspectrum}
\end{figure}

Figure~\ref{fig:energyspectrum} shows energy spectra of the single-hit and the multiple-hit events using the ES parameter cut. These spectra are compared with those of the COSINE-100 nominal event selections. DAMA/LIBRA's event selection introduces an excess below 2\,keV compared with the nominal COSINE-100 event selection. These events were categorized as PMT-induced noise in the nominal COSINE-100 event selection~\cite{COSINE-100:2020wrv,COSINE-100:2021xqn}. In the DAMA/LIBRA-phase2 energy spectrum, they also find a mild increase of the event rate below 2\,keV~\cite{Bernabei:2018yyw}, although these excess events were claimed as possible dark matter interactions~\cite{Bernabei:2020mon}. However, a possibility of remnants from PMT-induced noise for excess events of the DAMA/LIBRA data is not fully excluded.

Because DAMA/LIBRA's analysis always used residual spectra of the event rate by subtracting average backgrounds for the modulation fit, we first model the time-dependent background. Here we use two different approaches to account for the time-dependent backgrounds. The first model uses an exponential function to describe the time-dependent background that has been used for the initial annual modulation studies in COSINE-100~\cite{Adhikari:2019off} and ANAIS-112~\cite{Amare:2019jul}. The second model uses the yearly averaged rate as in the DAMA/LIBRA experiment~\cite{Bernabei:2018yyw}. 
		After subtracting the background, the residual rates are fitted with a sinusoidal function, 
		\begin{equation}
				R(t) = S_{m}\textrm{cos}\frac{2\pi(t-t_0)}{T},
				\label{eq:cosine}
		\end{equation}
		where $R(t)$ is the residual event rate as a function of time, $S_{m}$ is the modulation amplitude, $t_0$ is a phase, and $T$ is a period. 
		In the typical dark matter interaction assuming the standard halo model, $t_0$ and $T$ are expected to be June 2$^{\text{nd}}$ , and 365.25\,days (1\,year), respectively. 

\begin{figure}[!htb]
  \begin{center}
    \includegraphics[width=1.0\columnwidth]{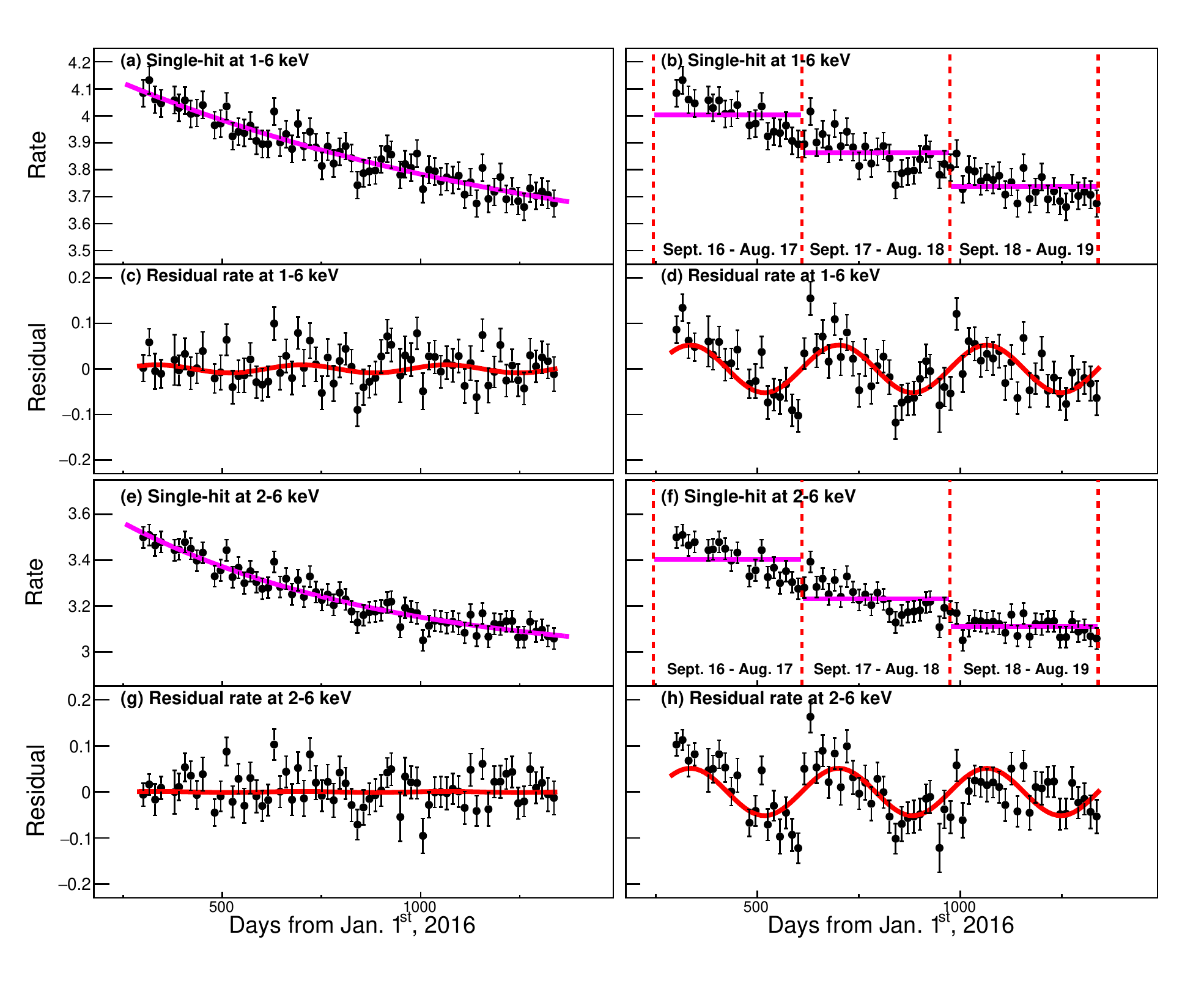} 
  \end{center}
  \caption{ 
   {\bf Single-hit event rates in the unit of counts/keV/kg/day as a function of time. }
	  The top four panels present time-dependent event rates as well as the residual rates in the single-hit 1--6\,keV regions with 15\,days bin. Here, the event rates are averaged for the five crystals with weights from uncertainties in each 15\,days bin size. Purple solid lines present background modeling with the single exponential (a) and the yearly averaged DAMA-like method (b). 
	Residual spectra for the single exponential model (c) and the DAMA-like model (d) are fitted with the sinusoidal function (red solid lines). Same for 2--6\,keV in the bottom four panels. Strong annual modulations are observed using the DAMA-like method while the result using the single-exponential models are consistent with no observed modulation.
}
  \label{fig:cosinerate_single}
\end{figure}

		Figure~\ref{fig:cosinerate_single} shows 1--6\,keV single-hit data and the time-dependent background model  with the single-exponential background model (a). 
		The residual spectrum (c) is fitted with the sinusoidal function to obtain the modulation amplitude $S_{m}$ = 0.0048$\pm$0.0055 counts/kg/keV/day. A similar procedure for the 2--6\,keV single-hit events ((e) and (g))  obtains $S_{m}$ = 0.0041$\pm$0.0056 counts/kg/keV/day.
		Even though the event selection only using the ES parameter contains significant PMT-induced noise events in the COSINE-100 data, the fitted results are consistent with no modulation when we model the time-dependent background using the single exponential function.

\begin{figure}[!htb]
  \begin{center}
    \includegraphics[width=1.0\columnwidth]{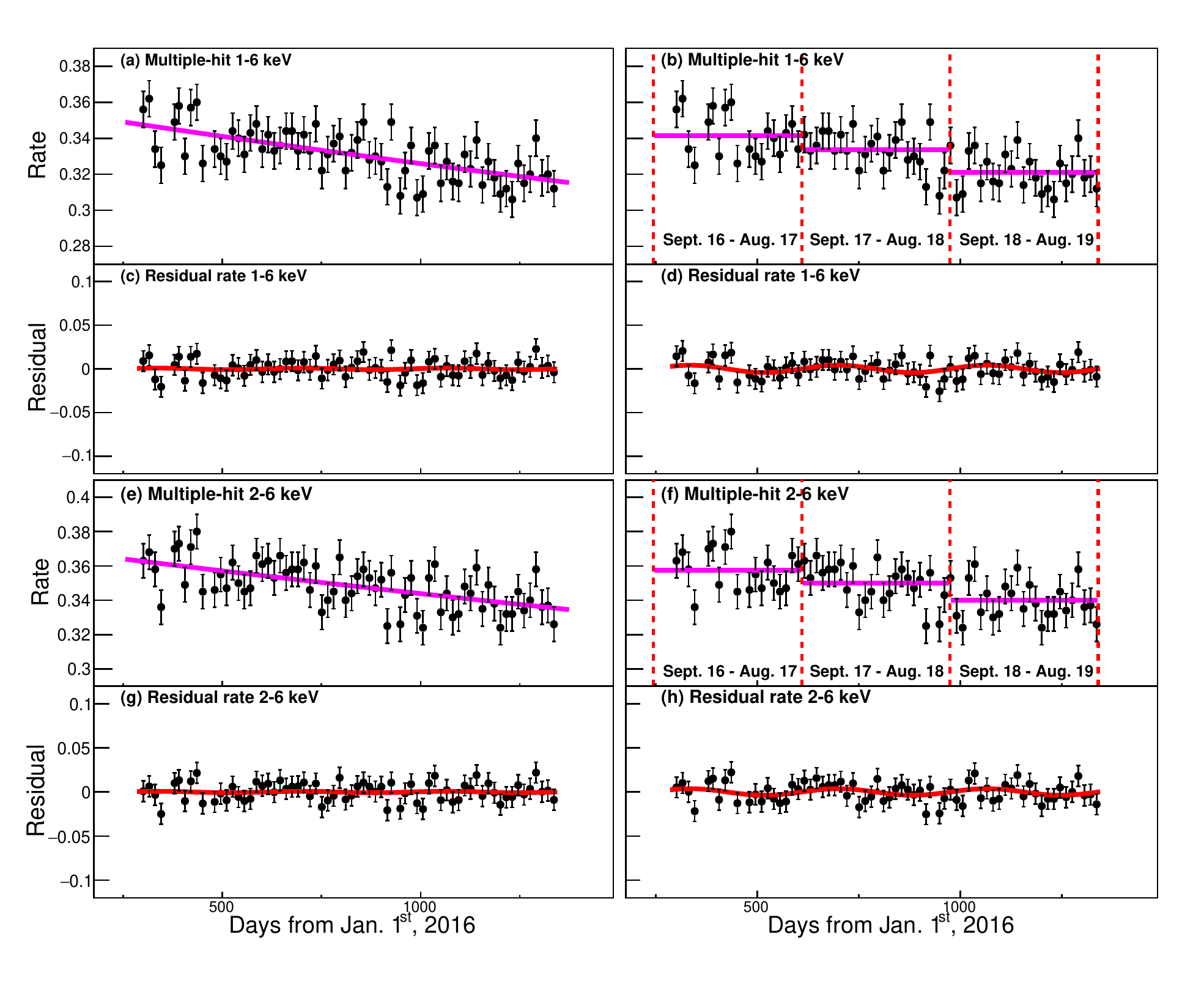} 
  \end{center}
  \caption{ 
   {\bf Multiple-hit event rates in the unit of counts/keV/kg/day as a function of time.}
The top four panels present time-dependent events rates as well as the residual rates in the multiple-hit 1--6\,keV regions with 15\,days bin size. Here, the event rates are averaged for the five crystals with weights from uncertainties in each 15\,days bin size. Purple solid lines present background modeling with the single exponential (a) and with the yearly averaged DAMA-like method (b). 
	Residual spectra for the single exponential model (c) and the DAMA-like model (d) are fitted with the sinusoidal function (red solid lines). The bottom four panels show the results for the 2-6 keV region which use the same methods. In the multiple-hit events, no strong modulations from both methods are observed. 
}
	 \label{fig:cosinerate_multiple}
\end{figure}

The same 1--6\,keV single-hit data are modeled with the DAMA-like year average method (b). 	
Considering 13 year cycles of DAMA/LIBRA shown in Table~\ref{tab:damayear}, we divide the COSINE-100 data into three year cycles presented in Table~\ref{tab:cycle}. The vertical lines in Fig.~\ref{fig:cosinerate_single} (b) represent the start and end of each year cycle. 
		The residual rates from this model are shown in Fig.~\ref{fig:cosinerate_single} (d) with the annual modulation fit. 
		Here we obtain a significant modulation amplitude $S_{m}$=$-$0.044$\pm$0.006\,counts/kg/keV/day, which has about 7\,$\sigma$ significance. 
		Similarly, in the 2--6\,keV energy region ((f) and (h)), $S_{m}$=$-$0.046$\pm$0.006\,counts/kg/keV/day is obtained. 
		The negative sign of $S_{m}$ indicates an opposite phase compared to DAMA/LIBRA and to the predicted phase from the WIMP dark matter model. 
		Because the COSINE-100 data have time-dependent backgrounds from cosmogenically activated nuclides and $^{210}$Pb, event rates are clearly decreasing as a function of time. 
		Simple yearly averages provide a bias on the modulation fit due to mismodeling of the time-dependent background. 
		Multiple-hit events in both 1--6\,keV and 2--6\,keV regions do not show significant modulation behaviors even though we apply the DAMA-like method as one can see in Fig.~\ref{fig:cosinerate_multiple}. 
This is because the multi-hit requirement removes the majority of radioisotopes that have high enough activities and short enough half-lives to appreciably affect the multi-hit event rate.
		Table~\ref{tab:resultNumbers} summarizes the fit results for the single-hit 1--6, 2--6\,keV events and comparisons of results from COSINE-100 (nominal analysis)~\cite{COSINE-100:2021zqh}, ANAIS-112~\cite{ANAIS2021}, and DAMA/LIBRA~\cite{Bernabei:2018yyw}.

\begin{table}[]
	\centering
	\caption{
			{\bf DAMA/LIBRA annual cycles}
	Thirteen annual cycles used by DAMA/LIBRA are obtained from Refs.~\cite{Bernabei2008yi,Bernabei:2013xsa,Bernabei:2018yyw}.}
	\begin{tabular}{cccccc}
		\hline
		\hline
		Cycle & Date period & Exposure (kg $\times$ day)\\
		\hline
		1&Sept/9, 2003 -- Jul/21, 2004 & 51,405\\
		2&Jul/21, 2004 -- Oct/28, 2005 & 52,597\\
		3&Oct/28, 2005 -- Jul/18, 2006 & 39,445\\
		4&Jul/19, 2006 -- Jul/17, 2007 & 49,377\\
		5&Jul/17, 2007 -- Aug/29, 2008 & 66,105\\
		6&Nov/12, 2008 -- Sept/1, 2009 & 58,768\\
		\hline
		7&Dec/23, 2010 -- Sept/9, 2011 & Commissioning\\
		8&Nov/2, 2011 -- Sept/11, 2012 & 62,917\\
		9&Oct/8, 2012 -- Sept/2, 2013 & 60,586\\
		10&Sept/8, 2013 -- Sept/1, 2014 & 73,792\\
		11&Sept/1, 2014 -- Sept/9, 2015 & 71,180\\
		12&Sept/10, 2015 -- Aug/24, 2016 & 67,527\\
		13&Sept/7, 2016 -- Sept/25, 2017 & 75,135\\
		\hline
		\hline
	\end{tabular}
	\label{tab:damayear}
\end{table}

\begin{table}[]
	\centering
	\caption{
			{\bf Three year cycles of COSINE-100 data for the DAMA-like method~\cite{Bernabei:2013xsa,Bernabei:2018yyw}.}
	}
	\begin{tabular}{lccccc}
		\hline
		\hline
		Cycle & Date period & Exposure (kg $\times$ day)\\
		\hline
		1 & Oct/20, 2016 -- Sept/16, 2017 & 20,323\\
		2 & Sept/17, 2017 -- Sept/26, 2018 & 22,963\\
		3 & Sept/27, 2018 -- Sept/20, 2019 & 22,042\\
		\hline
		\hline
	\end{tabular}
	\label{tab:cycle}
\end{table}

\begin{table}[]
	\centering
	\caption{ 
			{\bf Annual modulation amplitudes from various experiments.}
			The amplitudes of the annual modulation fits using the DAMA-like method to the COSINE-100 3\,years data (this work) are compared with results from DAMA/LIBRA~\cite{Bernabei:2013xsa,Bernabei:2018yyw}, COSINE-100~\cite{COSINE-100:2021zqh}, and ANAIS-112~\cite{ANAIS2021} in both 1--6\,keV and 2--6\,keV regions. 
	}
	\begin{tabular}{lcccc}
		\hline
		\hline
		 counts/kg/keV/day & 1--6\,keV  & 2--6\,keV  \\
		\hline
		This work       & -0.0441$\pm$0.0057 &  -0.0456$\pm$0.0056 \\
		DAMA/LIBRA      & 0.0105$\pm$0.0011  &  0.0095$\pm$0.0008 \\
		COSINE-100      & 0.0067$\pm$0.0042  &  0.0050$\pm$0.0047 \\
		ANAIS-112       & -0.0034$\pm$0.0042 &  0.0003$\pm$0.0037 \\
		\hline
		\hline
		
	\end{tabular}
	\label{tab:resultNumbers}
\end{table}

\begin{figure}[!htb]
  \begin{center}
    \includegraphics[width=1.0\columnwidth]{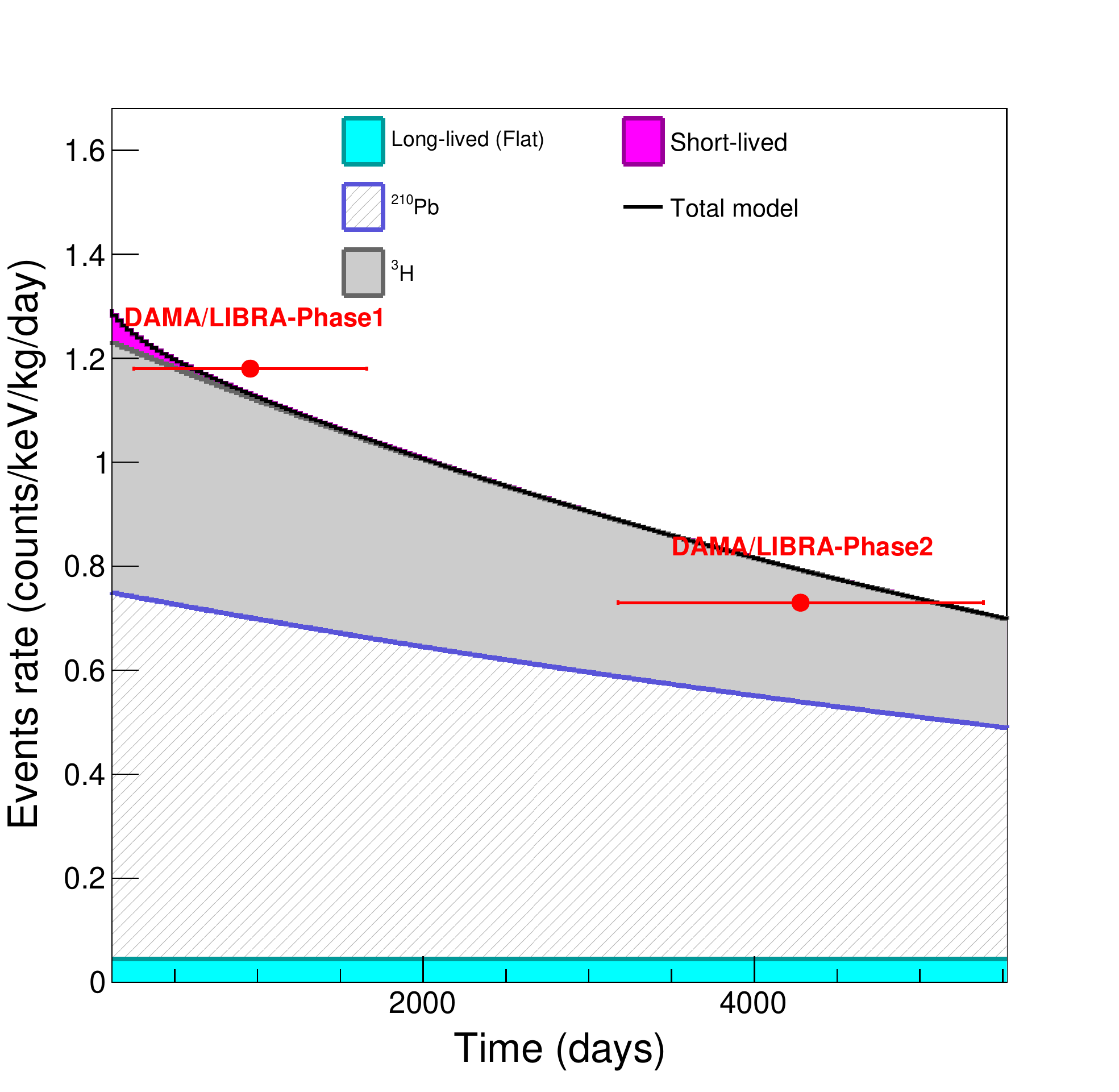} 
  \end{center}
  \caption{ 
    {\bf Background model for DAMA/LIBRA using the COSINE-100 background compositions.}
		Background compositions of the COSINE-100's crystal (crystal 6) in the 2--6\,keV single-hit regions are scaled to have 1.3\,counts/keV/kg/day at the beginning of DAMA/LIBRA-phase1. Here dominant background contributions are $^{210}$Pb (half-life of 8,140\,days) and $^{3}$H (half-life of 4,494 days). The total model (black solid line) is compared with the averaged rate in the 2--6\,keV region from the initial 4\,years of the DAMA/LIBRA-phase1~\cite{Bernabei2008yi} and 6\,years of the DAMA/LIBRA-phase2~\cite{Bernabei:2018yyw}. 
		Due to decays of the time-dependent backgrounds, this model presents a decreasing rate as a function of time. This model describes the rate of decrease obtained from the DAMA/LIBRA-phase1 to DAMA/LIBRA-phase2 denoted in the plot. 
}
  \label{fig:damamodel}
\end{figure}

   Following up the observation of the significant negative modulation of the COSINE-100 data using the DAMA-like method, 
	 we perform simulation studies about the DAMA/LIBRA's time-dependent background. 
	 Although the DAMA/LIBRA collaboration claimed no time-dependent background in their data~\cite{Bernabei:2020mon}, a clear decrease of the event rate in the 2--6\,keV range from DAMA/LIBRA-phase1~\cite{Bernabei2008yi} and DAMA/LIBRA-phase2~\cite{Bernabei:2018yyw} is observed, approximately from 1.2 to 0.7\,counts/kg/keV/day. 
	 The main update from DAMA/LIBRA-phase1 to DAMA/LIBRA-phase2 was the replacement of PMTs from ET~(Electron tubes) Enterprises to Hamamatsu Photonics for the high quantum efficiency~\cite{Bernabei_2012} but using the same NaI(Tl) crystals. 
	The observed decrease in backgrounds between phases is likely not due to the use of different PMTs, as there is a 10\,cm quartz light guide between PMTs and crystals. 
	 Background contribution from the PMT's radioisotopes in the COSINE-100 data, which used PMTs from the Hamamatsu photonics, was less than 0.05\,counts/kg/keV/day without the quartz block~\cite{cosinebg2}. 
	 We also noted that similar radiopurities were reported Electron tubes and Hamamatsu PMTs~\cite{Bernabei_2012}.
	 Therefore, one can suspect that the background rate decreasing from phase1 to phase2 was indeed due to decays of short-lived radioisotopes such as $^{3}$H and $^{210}$Pb that were reported by COSINE-100~\cite{deSouza:2019hpk,cosinebg2} and ANAIS-112~\cite{Villar2018aab}.
	 In this simulation study, we assume that DAMA/LIBRA's crystals have the same background composition as crystal 6 of COSINE-100, which has the lowest background among the COSINE-100 crystals.
	 The total average rate in the COSINE-100 crystal data is 2.5 times higher than total DAMA/LIBRA rate averaged over phase1 and phase2. We simulate DAMA/LIBRA data by scaling each COSINE-100 background component by the ratio of the average rate of 2--6\,keV single-hit events to be 1.3\,counts/kg/keV/day at the beginning of DAMA/LIBRA-phase1.
	 Because the dominant backgrounds in this region from the COSINE-100 crystal are from $^{3}$H and $^{210}$Pb, a decreased background rate is obtained as one can see in Fig.~\ref{fig:damamodel}. 
	 This background model plausibly describes the rate decrease from DAMA/LIBRA-phase1 to DAMA/LIBRA-phase2.

\begin{figure}[!htb]
  \begin{center}
    \includegraphics[width=1.0\columnwidth]{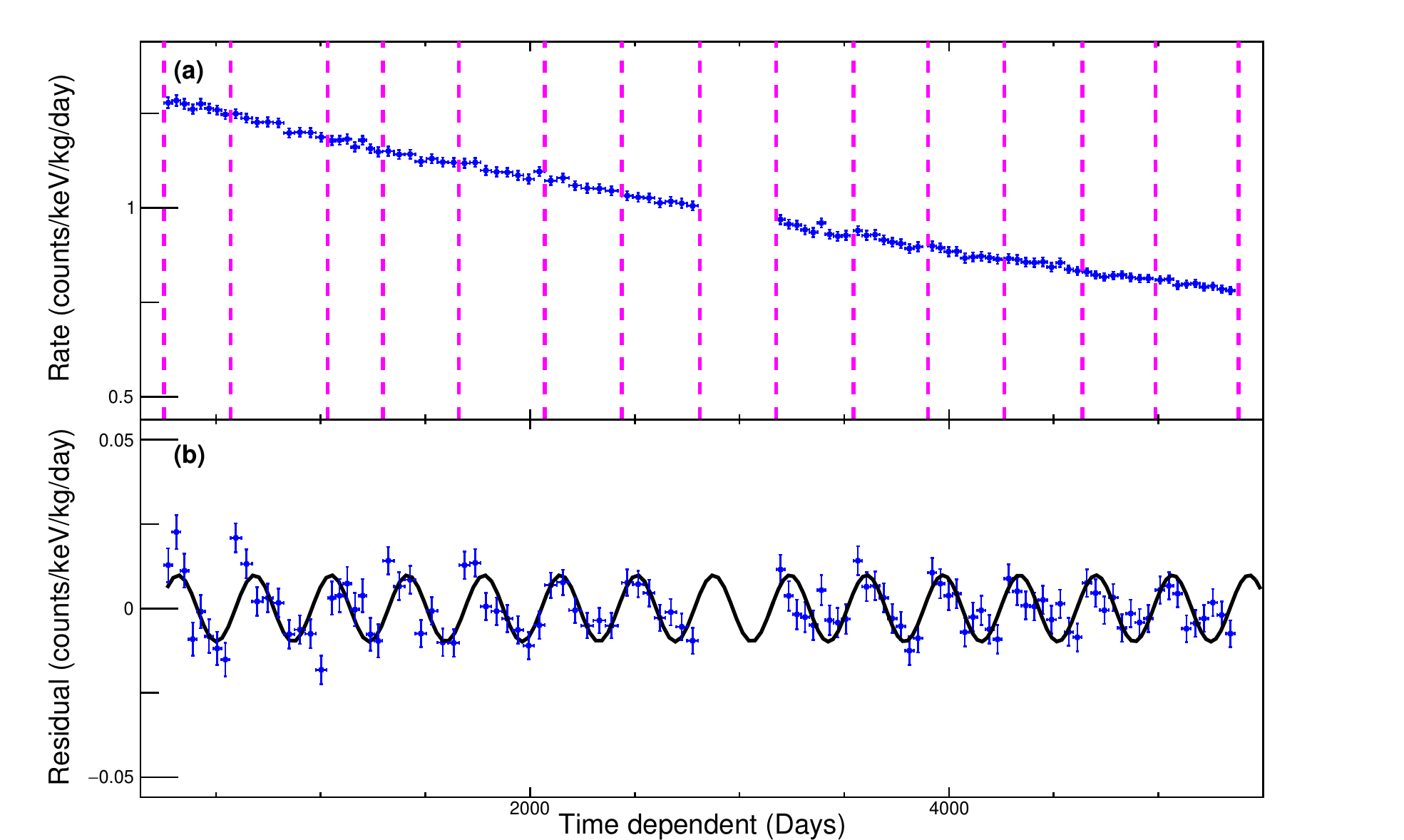}
  \end{center}
  \caption{
{\bf An example of a simulation of the the DAMA/LIBRA experiment in the single-hit 2--6\,keV region. }
(a) One sample of the simulated experiment of the time-dependent event rates for the DAMA/LIBRA assuming COSINE-100's background composition (points) is presented for 13\,year cycles. Vertical lines represent the start and end of each cycle used by the DAMA/LIBRA experiment. (b) Residual spectrum applying the DAMA-like method (points) is fitted with the sinusoidal function (solid line). Although no dark matter signals are inserted in this simulated data, strong modulation is observed from the DAMA-like method. 
}
\label{fig:pseudoresidual}
\end{figure}

	 Simulated data are generated from the aforementioned time-dependent background without the dark matter signals. 
	 Figure~\ref{fig:pseudoresidual} (a) presents an example of the simulated data for 13 year cycles of the DAMA/LIBRA experiment. 
	 Vertical lines present the start and end of one year cycle used by DAMA/LIBRA (see Table~\ref{tab:damayear}). 
	 The residual spectrum subtracting the yearly average rate in the DAMA-like method is presented in Fig.~\ref{fig:pseudoresidual} (b) with the modulation fit (solid line). 
	 Here we observe strong negative modulation of $S_{m}$ = $-0.0098~\pm~0.0008$ counts/kg/keV/day corresponding to approximately 12\,$\sigma$ significance.

\begin{figure}[!htb]
  \begin{center}
    \includegraphics[width=1.0\columnwidth]{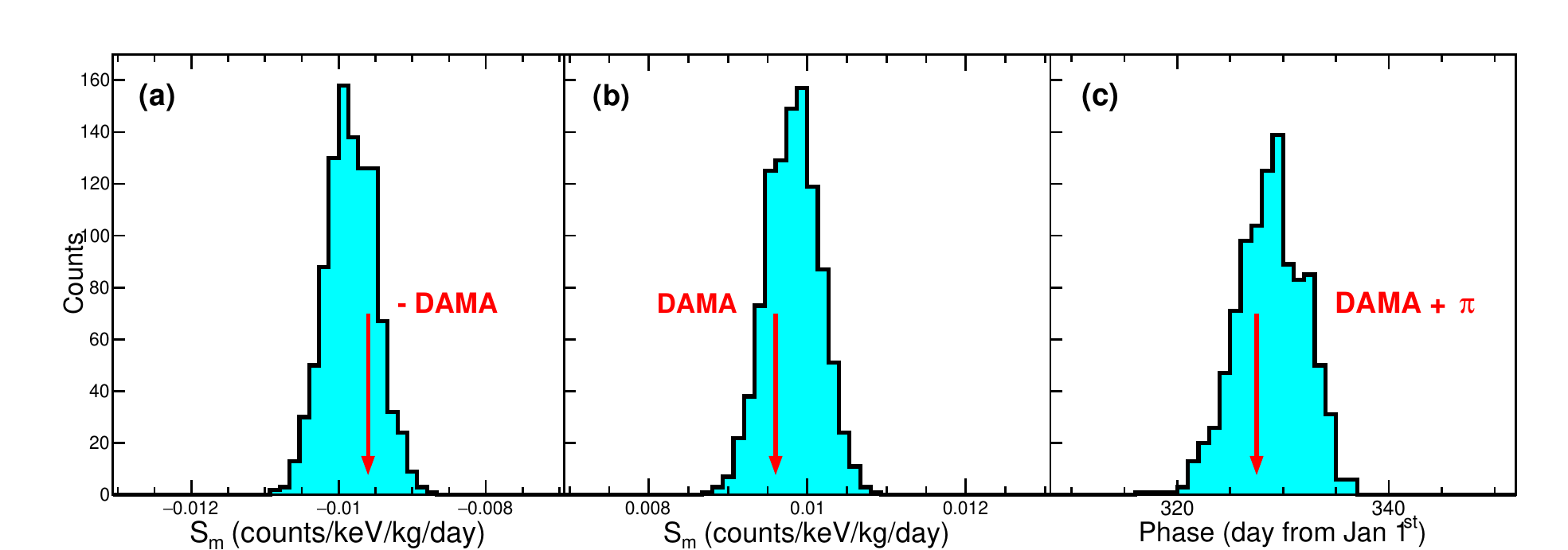} 
  \end{center}
  \caption{ 
    {\bf Results of 1,000 simulated experiments for the DAMA/LIBRA experiment. }
	We generate 1,000 simulated experiments with a background derived from COSINE-100 with an exposure and analysis method derived from the DAMA/LIBRA experiment without the dark matter signals. 
	(a) Results of the modulation amplitudes ($S_m$) using the phase-fixed fit are presented with DAMA/LIBRA's modulation amplitude. Although negative signs (opposite phase) are required, the simulated experiments obtain a modulation amplitude consistent with DAMA/LIBRA.
	Results for the phase-floated fits are shown for the modulation amplitude (b) and the modulation phase (c). Here we also observe consistent modulation amplitude 
but opposite phase by shifting the phase by $\pi$ between the simulated experiments and DAMA/LIBRA. 
}
  \label{fig:pseudo}
\end{figure}

	 We perform 1,000 simulated experiments and obtain the modulation amplitude distribution in Fig.~\ref{fig:pseudo} (a). Although negative signs of the modulation amplitudes are obtained, the magnitude of the modulation amplitude is consistent with that obtained by DAMA/LIBRA. 
	 We also perform the phase floated fits using the same simulated data. Figure~\ref{fig:pseudo} (b) and (c) show the modulation amplitudes and phases, respectively, compared with those of the DAMA/LIBRA experiment. Once again, results of similar modulation amplitude and opposite phase were observed.

	The direct detection of dark matter interactions is one of the main challenges of contemporary physics. 
	Because of its significance, any hints of dark matter evidence need to be precisely examined. 
	COSINE-100 continues to collect and analyze data to confirm or refute DAMA/LIBRA's hints at dark matter signals. 
	Alternative explanations of the signals, with not involving dark matter, are investigated with the same NaI(Tl) target materials from the COSINE-100 experiment. 
	We have observed PMT-induced noise contamination by adopting the selection criteria of DAMA/LIBRA. 
	The time-dependent background model using a single exponential decay function provided no modulation from the COSINE-100 data, although significant PMT-induced noise contained in the low-energy signal region was observed. 
	However, a DAMA-like yearly averaging method generates clear annual modulation signals because of mismodeling of the time-dependent background. 
	Furthermore, the DAMA-like method as applied to the simulated background-only data of DAMA/LIBRA, which are based on the time-dependent background in the COSINE-100 crystals, provide a consistent annual modulation signal of DAMA/LIBRA in the modulation amplitudes but the modulation phases are almost opposite to DAMA/LIBRA.
	Although the observed phase is opposite to the DAMA/LIBRA's observation, the consistent magnitude of the modulation amplitude may indicate interesting phenomenon hidden in DAMA/LIBRA's background subtraction procedure.

\bibliographystyle{PRTitle} 
\providecommand{\href}[2]{#2}\begingroup\raggedright\endgroup

\acknowledgments
We thank the Korea Hydro and Nuclear Power (KHNP) Company for providing underground laboratory space at Yangyang.
This work is supported by:  the Institute for Basic Science (IBS) under project code IBS-R016-A1 and NRF-2016R1A2B3008343, Republic of Korea;
NSF Grants No. PHY-1913742, DGE-1122492, WIPAC, the Wisconsin Alumni Research Foundation, United States;
STFC Grant ST/N000277/1 and ST/K001337/1, United Kingdom;
Grant No. 2021/06743-1 FAPESP, CAPES Finance Code 001, CNPq 131152/2020-3, Brazil.

\section*{Appendix}

\subsection*{Following DAMA/LIBRA's analysis method for the COSINE-100 data }

The shield structure of the COSINE-100 experiment is similar to that of the DAMA/LIBRA experiment. However, the COSINE-100 detector has additional active veto detectors including the outermost plastic scintillators for the muon veto~\cite{Prihtiadi:2017inr,Prihtiadi:2020yhz} and inner 2,200~L of liquid scintillator for the external or internal radiation veto~\cite{Adhikari:2020asl}. However, in this study we follow DAMA/LIBRA's event selection such that we do not use information from the plastic scintillators or the liquid scintillator, though they are still working as passive shields for external radioactivities. We defined two different categories of the events as the single-hit and the multiple-hit. The single-hit events correspond to any hit in a single NaI(Tl) crystal having no measurable energy in the other crystal. The multiple-hit events are defined in such a way that two or more NaI(Tl) crystals have measurable hits. 

In the data acquisition of DAMA/LIBRA, PMT signals from the two ends of the crystal are digitized by a waveform analyzer for a time window of 2048\,ns~\cite{BERNABEI2008297}. 
In the data analysis, only a 600\,ns time window integration of each pulse starting from the rising edge of each event was used to evaluate deposited energies. 
Low energies corresponding up to 100\,keV energy region were calibrated with external sources of $^{241}$Am (59.5\,keV) and $^{133}$Ba (30.4\,keV and 81.0\,keV) and with internal X-rays or $\gamma$s (3.2\,keV, 40.4\,keV, 67.3\,keV etc). A linear fits for those points was used for the energy calibration~\cite{BERNABEI2008297}. However, in the DAMA/LIBRA-phase2 they observed about 0.2\,keV shift for the tagged 3.2\,keV line and applied an additional correction in the low energy region between the software energy threshold and 15\,keV~\cite{Bernabei:2018yyw}. 

The data acquisition system of the COSINE-100 experiment took waveforms of events for 8,000\,ns time windows~\cite{Adhikari:2017esn,COSINE-100:2018rxe}. Nominal analysis used a 5,000\,ns time window starting from the rising edge of each event for the deposited energy. We used only internal X-rays or $\gamma$ lines for the energy calibration to avoid position dependencies from external sources. Nonproportional scintillation behavior of the NaI(Tl) crystals studied in Ref.~\cite{nonprop} was applied for energy calibration in the low-energy region~\cite{cosinebg2}. However, in this analysis, we follow the DAMA/LIBRA's method as closely as possible. Integrated charge in a 600\,ns time window is used for the deposited energy. A linear fit relating charge to energy is determined using the 59.5\,keV from external $^{241}$Am source and 3.2\,keV from internal $^{40}$K, providing an energy scale for the low-energy events. 

In the low-energy signal region below 10\,keV, PMT-induced noise events predominantly contribute to the single-hit physics data. The dominant noise has a fast decay time of less than 50\,ns compared with typical NaI(Tl) scintillation of about 250\,ns. The DAMA/LIBRA collaboration developed a good parameter based on ratios of slow ($X_1$) and fast ($X_2$) charges defined as following~\cite{BERNABEI2008297,Bernabei_2012}, 

\begin{eqnarray}
  X_{1} = \text{Charge (100 to 600 ns)}/\text{Charge (0 to 600 ns)},\\
  X_{2} = \text{Charge (0 to 50 ns)}/\text{Charge (0 to 600 ns)},
  \label{eq:waveform}
\end{eqnarray}
The typical PMT-induced noise deposited most of pulses in $X_2$ while scintillation events of the NaI(Tl) crystals deposited about 70\,\% scintillation charge in the $X_1$ area. For an effective rejection of PMT-induced noise events, DAMA/LIBRA defined an event selection (ES) parameter as following,  
\begin{eqnarray}
		\text{ES} = \frac{1-(X_{2}-X_{1})}{2}.
\label{eq:eventselection}
\end{eqnarray}

Typical signals have ES parameter around 0.8 while for the PMT-induced noise it is around 0.3, as one can see in Fig~\ref{fig:eventselection}. 
DAMA/LIBRA collected pure scintillation events using 59.5\,keV $\gamma$ rays from an $^{241}$Am source. 
DAMA/LIBRA determined the selection criteria as ES greater than 0.72 (0.85) in the energy region of 1--3\,keV (3--6\,keV)~\cite{BERNABEI2008297}.

Initially, we tried to use identical selection criteria for the COSINE-100 data, but the resulting selection efficiency was seen to differ significantly from that reported by DAMA/LIBRA~\cite{BERNABEI2008297,Bernabei_2012}.
This may be caused by slightly different scintillation characteristics of crystals due to different environmental conditions and different analysis methods used to determine the rising edge in the two experiments. 
Instead of the same ES parameter, we empirically develop cut criteria based on the selection efficiencies, making them similar between DAMA/LIBRA-phase2 and COSINE-100, as one can see in Fig.~\ref{fig:eventselection} and \ref{fig:efficiency}.

Because DAMA/LIBRA does not have the muon veto detectors, muon related events are not directly removed. Therefore, we tag high-energy events in each crystal requiring energies above 4\,MeV. 
We remove a 1\,s period of events in each crystal after the high energy event. 
In addition, data is monitored in two-hour periods which are occasionally removed when any large variation in the environmental or detector parameters are recorded. 

The low-energy spectra from the COSINE-100 data using the aforementioned calibration and event selection are presented in Fig.~\ref{fig:energyspectrum} and compared with the energy spectra from the nominal COSINE-100 data analysis~\cite{cosinebg2}. Noticeably, we observe a significant increase of the event rate below 2\,keV following DAMA/LIBRA's event selection method. Those excess events are categorized as PMT-induced noise events by the typical COSINE-100 data analysis.

\subsection*{Fitting procedure}
We calculate the event rate for each NaI(Tl) detector binned in 15-day intervals. We then evaluate the detector livetime in each time bin and normalize the event rate based on its relative exposure. This process accounts for variations in exposure induced by both detector-off periods and data periods that are removed due to detector instability.

Although time-dependent event rates were never presented by DAMA/LIBRA, they claimed no time-dependent backgrounds in their data~\cite{Bernabei:2020mon}.
The time dependence is reported for the residuals. 
The published DAMA/LIBRA residuals were found by subtracting the time-averaged ROI event rate for each one-year cycle from the measured rate.
The DAMA/LIBRA cycles start every year around September, as summarized in Table~\ref{tab:damayear}~\cite{Bernabei2008yi,Bernabei:2013xsa,Bernabei:2018yyw}. In addition to DAMA/LIBRA's method of the yearly averaged rate, we try to model the time-dependent background using a single exponential function to check the method bias. 
The residual rates are fitted to the sinusoidal functions. 

The data from five detectors are fitted simultaneously. We perform the fit by minimizing the value of the computed $\chi^{2}$, defined as 
\begin{equation}
\chi^{2} = \sum \limits _{ij} \frac{(n_{ij}-R_{ij})^{2}}{\sigma_{ij}^{2}},
\end{equation}
where $n_{ij}$ is the residual rate for $i$th crystal and $j$th time bin, and $\sigma_{ij}$ is the corresponding uncertainty of the data. 
$R_{ij}$ is the expected modulation rate, 
\begin{equation}
		R_{ij} = S_{m}\textrm{cos}\frac{2\pi(t-t_0)}{T}.
\end{equation}

In the phase-fixed fit, $t_0$ is fixed to June 2$^{nd}$ with period $T$=365.25\,days. A phase-floated fit with no constraints on $t_0$ is also performed. 

\subsection*{Time dependent background in DAMA/LIBRA and COSINE-100}

DAMA/LIBRA has not explicitly reported their background in detail, although a few individual radioisotope contaminations were studied~\cite{BERNABEI2008297}. 
In a recent paper~\cite{Bernabei:2020mon}, they presented the single-hit 1--100\,keV energy spectrum for the first time (Fig. 20 in Ref.~\cite{Bernabei:2020mon}). 
They described the data with $^{129}$I, $^{40}$K, and $^{210}$Pb decays, with continuum (constant) background due to high-energy $\gamma$/$\beta$, and signals (possibly dark matter interactions). 
Here, $^{210}$Pb can contribute to the time-dependent background due to its half life of 22.3\,years. 
In their modeling, $^{210}$Pb contribution in the ROI is negligible, so an assumption of no time-dependent background seems to be legitimate, assuming the validity of this particular background model.

In the ROI, dominant background contributions are caused by internal $^{40}$K and the continuum background. 
The continuum background may be caused by internal contamination of $^{238}$U, $^{232}$Th or external PMT radioactivities such as $^{238}$U, $^{232}$Th, and $^{40}$K. 
However, a precise modeling of the NaI(Tl) crystals performed by ANAIS-112 and COSINE-100 did not show continuum background from such components~\cite{Amare:2018ndh,cosinebg2}. 
Those backgrounds actually have decreasing rates in the low energy region. In this case, other backgrounds with potential to make increasing rates at low energy may need to be considered, examples are $^{3}$H and $^{210}$Pb in the crystal surface~\cite{cosinebg2,Yu:2020ntl}.
When considered, a time-dependent event rate which decreases in time is natural and similar to what is observed in ANAIS-112~\cite{ANAIS2021} and COSINE-100~\cite{COSINE-100:2021zqh}. It is worthwhile to mention that the background event rate in the 2--6\,keV region in DAMA/LIBRA has decreased from DAMA/LIBRA-phase1 to DAMA/LIBRA-phase2 as indicated in Fig.~\ref{fig:damamodel}. 

For the simulated experiments of DAMA/LIBRA, we try to generate the time-dependent background describing DAMA/LIBRA's background behavior.
Because there is no explicit description of time-dependent backgrounds from DAMA/LIBRA, we simply consider the same background components as the COSINE-100 crystals. 
The approximately 2.5 times lower background is scaled from the COSINE-100 background using the same fractional compositions of different background components. Short-lived radioisotopes with half lives less than 100\,days were not included. 
Averaged rate of each time-dependent background component at the beginning of the experiment in the 2--6\,keV region is summarized in Table~\ref{tab:radioisotope}. 
Figure~\ref{fig:damamodel} shows the model of the time-dependent background in the 2--6\,keV region derived from the COSINE-100 data. Interestingly, this decreasing rate has an agreement with DAMA/LIBRA-phase1 and phase2 data in the averaged event rates. With this model, we generate simulated data and perform the fit for the year-subtracted residual rate as shown in Fig~\ref{fig:pseudoresidual}.

\begin{table}[]
	\centering
	\caption{
			{\bf  Time-dependent background contributions in the NaI(Tl) crystals.} 
	  The model of DAMA/LIBRA background, assuming COSINE-100's background composition is summarized with half lifes and initial background contributions in the 2--6\,keV single-hit events. 
}
	\begin{tabular}{lccccc}
		\hline
		\hline
		Isotopes &  Half-lives(day) & Counts/day/kg/keV\\
		\hline
		$^{238}$U, $^{232}$Th, $^{40}$K~~~~& $>$10$^{10}$ & 0.043\\
		$^{210}$Pb& 8,140 & 0.687\\
		$^{3}$H& 4,494 & 0.474\\
		$^{113}$Sn& 115.1 & 0.055\\
		$^{109}$Cd& 462 & 0.025\\
		$^{121m}$Te& 164.2 & 0.004\\
		$^{127m}$Te& 106.1 & 0.011\\
		\hline
		\hline
	\end{tabular}
	\label{tab:radioisotope}
\end{table}
\clearpage
\end{document}